\documentclass[aps,prl,twocolumn,groupedaddress,showpacs,floatfix]{revtex4}
\usepackage{graphicx}
\begin{document}
\title{The stellar ($n,\gamma$) cross section of $^{62}$Ni}

\author {
H. Nassar$^1$, 
M. Paul$^1$\footnote{{To whom correspondence should be addressed,
email address: paul@vms.huji.ac.il}},
I. Ahmad$^2$,
D. Berkovits$^3$,
M. Bettan$^3$, 
P. Collon$^4$,
S. Dababneh$^5$,
S. Ghelberg$^1$,
J.P. Greene$^2$,
A. Heger$^6$,
M. Heil$^5$,
D.J. Henderson$^2$,
C.L. Jiang$^2$,
F. K\"appeler$^5$,
H. Koivisto$^7$,
S. O'Brien$^4$,
R.C. Pardo$^2$,
N. Patronis$^8$,
T. Pennington$^2$\footnote{{Deceased}},
R. Plag$^5$,
K.E. Rehm$^2$,
R. Reifarth$^6$,
R. Scott$^2$,
S. Sinha$^2$,
X. Tang$^2$,
R. Vondrasek$^2$
}

\affiliation{$^1$ Racah Institute of Physics,
Hebrew University, Jerusalem, Israel, 91904}
\affiliation{$^2$ Argonne National Laboratory, Argonne, IL 60439, USA}
\affiliation{$^3$ Soreq Nuclear Research Center, Yavne, Israel 81800}
\affiliation{$^4$  Physics Department, University of Notre Dame,
Notre Dame, IN 46556, USA}
\affiliation{$^5$ Forschungszentrum Karlsruhe Institut f\"{u}r Kernphysik, 
PF 3640, 76201 Karlsruhe, Germany}
\affiliation{$^6$ Los Alamos National Laboratory, Los Alamos, 
NM 87545, USA}
\affiliation{$^7$ Department of Physics, 
University of Jyv\"{a}skyl\"{a}, FIN-40351, Finland}
\affiliation{$^8$ Department of Physics, The University of Ioannina, 
45110 Ioannina, Greece}

\date{\today}

\begin{abstract}
The $^{62}$Ni($n,\gamma$)$^{63}$Ni(t$_{1/2}$=100$\pm$2 yrs)
reaction plays an important role 
in the control of the flow
path of the slow neutron-capture ({\it s-}) 
nucleosynthesis process. 
We have measured for
the first time the total cross section of this reaction 
for a quasi-Maxwellian ($kT\!=\!25$~keV) neutron flux. 
The measurement was performed by 
fast-neutron activation, combined with
accelerator mass spectrometry to detect directly
the $^{63}$Ni product nuclei.
The experimental value of  $28.4\pm 2.8$ mb,
fairly consistent with a recent 
calculation, 
affects the calculated net yield of $^{62}$Ni itself 
and the
whole distribution of nuclei with $62<\!A<\!90$ produced by the 
weak {\it s}-process in massive stars.

\end{abstract}


\maketitle

 The quest for experimentally determined cross sections of stellar
nucleosynthesis reactions becomes more focused with the refinement
of theoretical models and calculation techniques. 
Above A$=\!60$ and  except for rare cases, nuclei
are created by captures of free neutrons 
in two main mechanisms,
the slow ({\it s-})process and rapid ({\it r-})process
\cite{sne:03}.
The path of the 
weak {\it s}-process component, related to 
helium burning in massive stars of 10 to 25~$M_{\odot}$ 
(M$_{\odot}$ denotes one solar mass)\cite{rai:93},
starts with the major $^{56}$Fe seed nucleus 
and, proceeding through the neutron-rich Ni region,
dominates the synthesis of nuclei with masses between
60 and 90.
The particular
importance of the cross section of neutron capture reactions 
in the Ni region
has been emphasized by Rauscher {\it et al.} \cite{rau:02}
who suggest
that inaccurate cross section values may be responsible for 
the overestimate of neutron-rich Ni isotope abundances
(compared to solar abundances)
in nucleosynthesis calculations. The role of the 
$^{62}$Ni($n,\gamma $)$^{63}$Ni reaction
is critical in this respect because it affects 
the entire
weak {\it s}-process flow starting from $^{56}$Fe. 
In that regime, 
the neutron exposure is not strong enough to drive the system to
the so-called local equilibrium and
the neutron capture rate of a nucleus 
such as $^{62}$Ni, 
will affect not only the 
calculated abundance of this nucleus itself, 
but also that of all following ones.
The experimental information on the $^{62}$Ni($n,\gamma $)$^{63}$Ni
reaction cross section  at keV energies before this work
relied on neutron 
time-of-flight (TOF) measurements
aimed at determining resonance parameters \cite{bee:74,bee:75}. 
The more recent values of the  
Maxwellian-averaged capture cross section 
(MACS, see definition in \cite{bee:92})
at $kT\!=\!30$~keV, evaluated from
these data and from the thermal (2200 m/s) neutron-capture
cross section \cite{sim:70}, 
are 35.5$\pm4$~mb  (\cite{bao:87},
\cite{bee:92})
and 12.5$\pm$4 mb \cite{bao:00}. 
The decrease by a factor of $\sim$3
in \cite{bao:00} results from
the subtraction of the contribution 
of a sub-threshold resonance, prior to
extrapolation to higher energies.
Moreover, as later emphasized in ref.\cite{rau:02a}, none of these 
evaluations include a possibly 
significant direct-capture (DC) component. In ref.\cite{rau:02a}, the
DC $s-$ and $p-$wave contributions were calculated and added to
the resonant component to evaluate the 
$^{62}$Ni($n,\gamma $)$^{63}$Ni MACS up to $kT\!\!=\!500$~keV;
the MACS at $kT\!=\!30$~keV is estimated
as 35$\pm5$~mb. We report here the first experimental
determination of the total
$^{62}$Ni($n,\gamma $)$^{63}$Ni cross section for quasi-Maxwellian
neutrons ($kT\!=\!25$~keV). The measurement was performed by 
a fast-neutron activation coupled for the first time with  
accelerator mass spectrometry (AMS). 
Detection of the
decay radiation of the 
long-lived product
$^{63}$Ni($t_{1/2}=100\pm2$ yrs, $\beta^-$ 
endpoint energy of 66~keV and no $\gamma$) is 
impractical 
and is replaced in the AMS technique (see \cite{tun:98})
by the 
counting of $^{63}$Ni ions, 
unambiguously
identified after acceleration to several MeV/{\it u}.
The MACS 
$\sigma_V=\langle\sigma v\rangle/v_T $ is extracted
from the AMS determination of the
isotopic ratio 
in the activated
material, using the integrated neutron flux 
$\langle\phi t\rangle$ monitored during activation.

A 36.8 mg Ni metal sample, enriched to 95\% in $^{62}$Ni,
was activated at the 
Karlsruhe 3.7~MV Van de Graaff accelerator using 
a fast-neutron beam 
obtained from the $^7$Li($p,n$)$^7$Be reaction \cite{bee:80,rat:88}
at a proton energy of 1911~keV,
just above threshold.
In order to ease the handling of the small sample, the metal powder
was dispersed in a camphor matrix 
($>95$\% C$_{10}$H$_{16}$O, 53 mg)
and pressed into a pellet; 
after activation, the metal
was recovered (36.2 mg) 
by subliming the camphor under vacuum.  
In our conditions,
all neutrons are emitted in a forward cone
of 120$^\circ$ opening angle and the angle-integrated spectrum provides
a good approximation to a thermal spectrum 
at $kT\!=\!25\pm 0.5$~keV \cite{bee:80,rat:88}
with an upper cut-off energy of 106 keV.
Any moderation of the neutron spectrum due to the
camphor matrix is negligible in the conditions
of the experiment.
Before each run, 
the accelerator was operated in pulsed mode with a repetition
rate of 1~MHz and a pulse width of 10~ns. This allowed the measurement of 
the neutron energy spectra via TOF using a $^6$Li 
glass monitor located 1~m downstream of the  $^7$Li target.
The proton energy is verified via the cut-off energy in the spectrum.
During activations, the accelerator was run in
direct-current mode with beam intensities around 100~$\mu$A. 
The time-integrated
neutron flux $\langle\phi_{exp} t\rangle = 
(3.68\pm 0.14)\times 10^{15}$ cm$^{-2}$,
collected over 17.4 days,
was continuously monitored by two
Au foils
of the same size and
sandwiching the sample
and determined using the experimental value $586\pm8$~mb \cite{rat:88}
of the
capture cross section of $^{197}$Au for neutrons with 
quasi-stellar ($kT\!=\!25$~keV) energy distribution.
The $^6$Li glass monitor recorded the 
neutron yield at regular intervals and the resulting flux history was 
used to evaluate the corrections for decay of the Au monitors
during activation (see \cite{bee:80} for details).

To serve later as  calibration in the AMS measurement,
a 321.4 mg high-purity $^{nat}$Ni metal powder sample 
was activated for 30 s 
at the rabbit facility of the Soreq  
nuclear reactor in a thermal neutron flux. 
Au monitors (bare or shielded by a Cd tube), separately activated in the 
same conditions, yielded a Cd ratio of
10.7$\pm$2.3. After 
neutron irradiation, the $^{65}$Ni(2.5 hrs)
$\gamma$ activity produced in the $^{nat}$Ni sample 
was measured  in a 100~cm$^3$  
Ge(Li) detector at a distance of  7.7~cm 
in order to monitor the integrated neutron flux. 
The efficiency of the
detector was measured, using a vial which contained 
0.1413 g of a $^{152}$Eu calibrated solution (6.44$\pm$0.17 kBq) and
reproduced 
very closely 
the geometry of the $^{nat}$Ni activated sample. 
The efficiency was determined for the three
major $\gamma$ lines in $^{65}$Ni decay (366.3, 1115.5 and 1481.8~keV)
by fitting the yields of $\gamma$ lines in the $^{152}$Eu decay
at 344.3, 1112.1 and 1408.0~keV, very close in energy to those of $^{65}$Ni
and including minor corrections for summing effects.
From the measured $^{65}$Ni activity
(corresponding to $(4.58\pm 0.12)\times 10^9$ (1$\sigma$)
$^{65}$Ni atoms produced in the activation),
the thermal-neutron capture and resonance integral of $^{64}$Ni
($\sigma_{\gamma}= 1.52 \pm 0.03$ b and I$_{\gamma}=0.98 \pm 0.3$ b
\cite{gry:78}, respectively)
and those of $^{62}$Ni $(\sigma_{\gamma}= 14.2 \pm 0.3$ b, 
I$_{\gamma}=9.6 \pm 3.5$ b \cite{sim:70}), 
and the isotopic abundances of 
$^{nat}$Ni,
we determine a ratio 
$^{63}$Ni/$^{58}$Ni$=(7.52\pm 0.35)\times 10^{-11}$ (1$\sigma$)
in the reactor-activated $^{nat}$Ni sample.

The AMS detection of $^{63}$Ni was
\begin{figure}
\includegraphics[width=85mm,height=60mm]{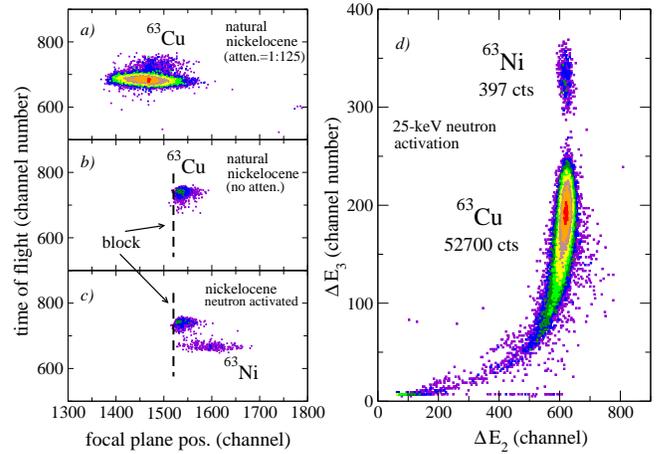}
\caption{\label{63Ni_spect} Gas-filled magnet separation of
the $^{63}$Cu-$^{63}$Ni isobaric pair :
{\it a-c)} Two-dimensional time-of-flight vs.
focal plane position spectra for: {\it a,b)}
a $^{nat}$Ni (nickelocene) sample in the ECR ion source;
{\it c)} activated $^{nat}$Ni. In {\it a)}, the ion beam
was attenuated by a factor of 125 to permit the counting
of the full $^{63}$Cu isobaric group.
In {\it b)} and {\it c)}, most of the  $^{63}$Cu group is
blocked before the detector by a movable shield;
{\it d)} Identification spectrum of $^{63}$Ni ions in the detector
blocked as in {\it c)}, measured for the fast-neutron
activated sample. The {\it x} ({\it y}) axis represents the energy loss
measured in the second (third) anode of the focal-plane ionization chamber.}
\end{figure}
performed at Argonne National Laboratory,
using the Electron Cyclotron Resonance (ECR) ion source \cite{sch:98s}, 
acceleration of $^{63}$Ni$^{15+}$ ions
to 9.2 MeV/{\it u} with the ATLAS facility and the gas-filled
magnetic spectrograph technique \cite{pau:89}
to separate the extremely intense
stable isobaric $^{63}$Cu component. A preliminary experiment \cite{nas:03s}
had shown that feeding of Ni into the ECR plasma chamber in gas
phase from nickelocene (Ni(C$_5$H$_5$)$_2$) organo-metallic compound 
reduces the $^{63}$Cu contaminant
by a factor of $\sim$100 compared
to solid material. A similar
gas-feeding technique was recently used for the detection of $^{63}$Ni 
by AMS using a Cs-sputter negative ion source \cite{str:03}.
For the present experiment, both the fast-neutron and the reactor-neutron
activated samples were chemically converted to nickelocene at the 
University of Jyv\"{a}skyl\"{a} \cite{koi:94}.
In order to ensure reliable yields in the synthesis process 
which required Ni sample masses above
$\sim$100 mg,
the fast-neutron activated sample was mixed 
with 101.3 mg of high-purity
$^{nat}$Ni before nickelocene synthesis.
The chemical synthesis
(involving dissolution of the metal in HCl, reaction with
NaC$_5$H$_5$, evaporation to dryness and sublimation of Ni(C$_5$H$_5$)$_2$),
ensured complete isotopic homogeneization
of the sample, 
resulting also in averaging
the fast-neutron activation 
over the whole angular range; 
244.6 mg and 668.5 mg 
nickelocene from the fast-neutron and reactor-activated samples 
respectively, were obtained. It is important to note that the efficiencies
of the chemical procedure or ion production and transport
do not affect the measurements of isotopic ratios, assuming
that mass fractionation effects therein are negligible.
The ATLAS accelerator system was entirely tuned with a stable beam 
of $^{84}$Kr$^{20+}$ prior to the AMS measurement of $^{63}$Ni$^{15+}$.
The two ions have nearly equal $m/q$ 
mass-to-charge ratios ($\delta (m/q)/(m/q)= 6.4\times 10^{-5}$)
and are assumed to be transported identically from the 
ion source to the detection system. The latter was a
large hybrid detector composed of a $(x,y)$ position-sensitive 
parallel-grid avalanche counter (measuring also the ion TOF
relative to the accelerator RF trigger) 
and a multi-anode ionization chamber 
\cite{pau:97a},
located in the focal plane of the gas-filled Enge split-pole magnetic 
spectrograph. 
The entrance aperture of the detector (48~cm$\times$10~cm)
ensured 100\% acceptance of the ions reaching the focal plane.
The spectrograph, filled with N$_2$ at 20 Torr, focused the
$^{63}$Ni and $^{63}$Cu ions  into groups $\sim 3$ cm 
wide (FWHM), separated from each other by
$\sim$ 5.5 cm (Fig.~\ref{63Ni_spect}{\it a-c}). 
This physical separation
allows one to block more than 95\% of the extremely intense 
$^{63}$Cu group ($\sim 7\times 10^4$ ions/s) 
and let the entire $^{63}$Ni group enter into
the detector for further discrimination. The latter is achieved by
multiple energy-loss measurements along the ion path in the 
ionization chamber (Fig.~\ref{63Ni_spect}{\it d}).  
\begin{figure}[top]
\includegraphics[width=80mm,height=51mm]{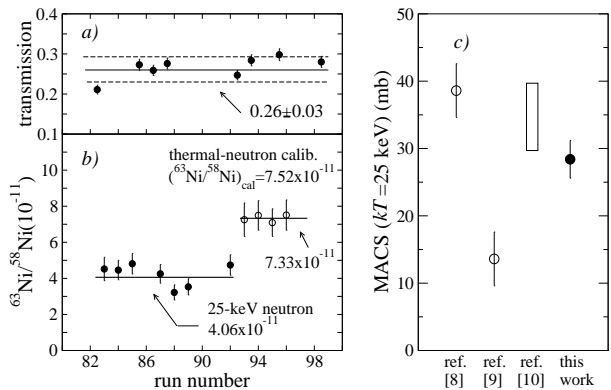}
\caption{\label{ams_results}{\it a)} Beam transmission through 
the accelerator during the experiment. 
The solid and dashed lines
are the unweighted mean and standard deviation
of the measured values respectively;
{\it b)} Repeated measurements of the 
$^{63}$Ni/$^{58}$Ni ratio for the fast-neutron activated sample
(solid dots) and the calibration thermal-neutron 
activated sample (open dots).The error bars include 
(in quadrature) statistical counting
error, standard deviation of the mean beam transmission ({\it a})
and an estimated error on the $^{58}$Ni$^{15+}$  
mean current during each run
(see text). The solid lines indicate the weighted mean
of the measurements;
{\it c)} Comparison of the present experimental
value of the total Maxwellian averaged 
$^{62}$Ni($n,\gamma$)$^{63}$Ni cross section ($kT\!=\!25$~keV)
(solid dot) with evaluations from
\cite{bao:87}, \cite{bao:00} (open dots) and of \cite{rau:02a} (open box).
The values from \cite{bao:87} and \cite{bao:00} 
were corrected for a thermal
energy of 25~keV, using the temperature dependence of \cite{bao:00},
in order to compare them with the experimental value. 
} 
\end{figure} 

The beam transmission through the accelerator 
was repeatedly monitored 
(Fig.~\ref{ams_results}{\it a}) during the experiment 
between sample measurements, by injecting a
small amount of isotopically enriched $^{84}$Kr gas into the ion source
and measuring the ion current at the injection line and before the
spectrograph. The value and random error adopted for the 
beam transmission 
were the unweighted mean and standard deviation of
the repeated measurements, respectively. 
The count rate of identified
$^{63}$Ni ions (Fig.~\ref{63Ni_spect}{\it d}), corrected for
beam transmission (Fig.~\ref{ams_results}{\it a}) 
and for a detection efficiency of
0.90 
(purely systematic) 
due to grid shadowing,
is divided by the ion current of the
stable $^{58}$Ni$^{15+}$ beam measured after mass analysis at the
accelerator injection line, to yield the $^{63}$Ni$^{15+}$/$^{58}$Ni$^{15+}$
isotopic ratio (Fig.~\ref{ams_results}{\it b}).
An additional correction factor of 0.90$\pm$0.04
in the $^{58}$Ni$^{15+}$ ion current measured in the injection line
was deemed necessary for the
fast-neutron activated sample, due to the near degeneracy of
$^{58}$Ni$^{15+}$ and $^{62}$Ni$^{16+}$ beams. This correction was
obtained by scanning the intensity of the main
Ni isotopes and charge states. It is negligible in the case
of the reactor-activated sample which has a
very small (natural) $^{62}$Ni abundance.
In order to reduce unknown systematic errors in
the determination of the isotopic ratios,
the latter were normalized to
the calibration sample (reactor-activated) by multiplying
by a factor 1.03$\pm$0.08 (ratio between
the values  obtained  for the reactor-activated sample in
$\gamma$ and AMS measurements, see Fig.~\ref{ams_results}{\it b}).
From the final determination in the fast-neutron
activated (diluted) sample
$^{63}$Ni/$^{58}$Ni = $(4.17\pm 0.37)\times 10^{-11}$
(weighted mean and standard deviation of the mean of repeated measurements),  
the number of $^{58}$Ni atoms in the sample and 
the number of $^{62}$Ni atoms in the initial $^{62}$Ni-enriched sample, 
we derive a ratio 
$r=^{63}$Ni/$^{62}$Ni= $(8.78\pm 0.79)\times 10^{-11}$.

We define the experimental capture cross
section as 
$\sigma_{exp}=
\int_{0}^{E_{co}} \sigma(E) \phi_{exp}(E)dE /\int_0^{E_{co}}\phi_{exp}(E)dE$,
where $\phi_{exp}(E)$ denotes the experimental 
neutron energy spectrum and E$_{co}$ its cut-off 
energy (E$_{co}$= 106~keV). 
$\sigma_{exp}$ is related to 
the measured isotopic ratio $r$  
by 
$\sigma_{exp}= r/\int_{0}^{E_{co}}\int\phi_{exp}(E,t) dt dE=
r/\langle \phi_{exp}t\rangle$, 
where we implicitely 
use the fact that $\phi_{exp}(E,t)$ 
varies only in intensity during the experiment,
its energy distribution $\phi_{exp}(E)$ staying unchanged.  
A correction factor $f_{\infty}$
must be applied to $\sigma_{exp}$ to account for
the $^{62}$Ni$(n,\gamma)$ yield at energies $E\!>\!E_{co}$.
$f_{\infty}$ was calculated as $1.05\pm0.01$, using for
$\sigma(E)$ expressions derived in  \cite{rau:02a} for
direct capture. This mechanism was shown in \cite{rau:02a}
to be dominant, as also confirmed 
experimentally by the present work (see below).
The $^{62}$Ni$(n,\gamma)^{63}$Ni MACS at  $kT\!=\!25$~keV,
$\sigma_V = \int \sigma v \phi (v) dv/v_T$, 
is finally given by $\sigma_V = (2/\sqrt \pi)f_{\infty}\sigma_{exp}$ 
\cite{bee:92, rat:88}. From our measured values,
we determine 
$\sigma_V(^{62}$Ni)= $28.4\pm 2.8$~mb. This value  
is compared in 
Fig. \ref{ams_results}{\it c}) with existing evaluations.
The recent calculation of \cite{rau:02a} including a 
direct-capture component is fairly consistent
with the experimental value.

\begin{figure}
\includegraphics[width=95mm,height=100mm]{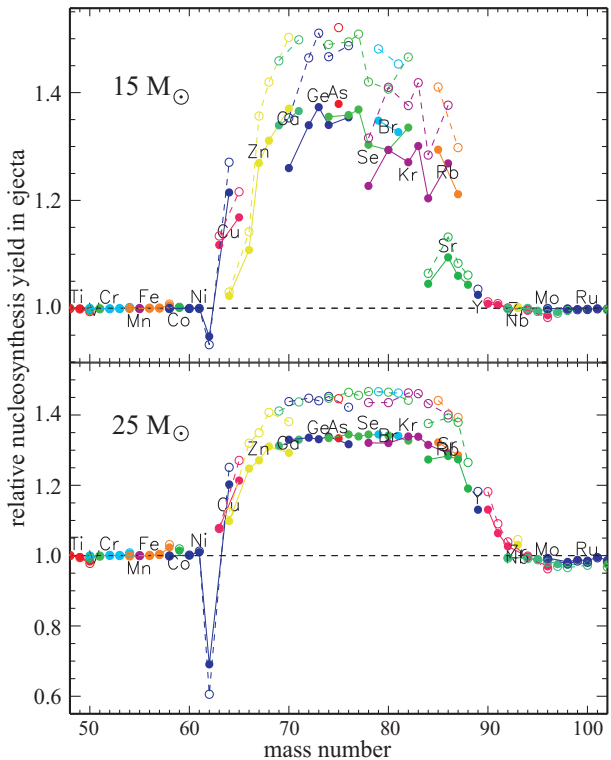}
\caption{\label{15_25_msol} Nucleosynthesis yields as
function of mass number calculated 
for different values of the $^{62}$Ni$(n,\gamma )^{63}$Ni
cross section at $kT\!=\!30$~keV
and normalized to the yields at a cross section of
12.5 mb \cite{rau:02}, 
for a 15~M$_{\odot}$ (top) and 25~M$_{\odot}$ (bottom) star.
Solid dots represent the ratios between the yields calculated using 
26.1~mb (this work, extrapolated from 28.4~mb at $kT\!=\!25$~keV) 
to that using 12.5~mb \cite{bao:00}.
Open dots represent the ratios between yields for 
35.5~mb 
\cite{bao:87, rau:02a} and 12.5~mb \cite{bao:00}. Solid (open) 
dots corresponding 
to isotopes of a given element are connected by solid (dashed) lines.}
\end{figure}

Figure \ref{15_25_msol} 
illustrates the results of stellar-evolution calculations 
when different values of  
$\sigma_V(^{62}$Ni) are used, relative 
to the results of models S15 and S25 \cite{rau:02} for a 
15~M$_{\odot}$ and 25~M$_{\odot}$ star respectively.
The calculations differ solely by the 
value adopted for
$\sigma_V(^{62}$Ni): 
28.4~mb ($kT\!=\!25$~keV; this
work),  35.5 mb ($kT\!=\!30$~keV; \cite{bao:87})
or 12.5 mb ($kT\!=\!30$~keV; \cite{bao:00,rau:02}) and assume
the same temperature dependence of $\sigma_V(^{62}$Ni) 
as in \cite{bao:00}.
The mass region $60\!<\!\!A\!<\!\!90$ exhibits the characteristic abundance
peak due
to the weak {\it s-}process in massive stars  \cite{rai:93}. 
These {\it s}-yields are
strongly
influenced by the change in $\sigma_V(^{62}\mbox{Ni})$.
The increase in this cross section 
from 12.5~mb to the present value of 28.4~mb
has two effects:
({\it i}) the overproduction of $^{62}$Ni itself \cite{rau:02}
is damped by the larger destruction probability,
as indicated by the dip in Fig. \ref{15_25_msol};
({\it ii}) its effect
of bottle neck in the reaction flow between the major seed nuclei
$^{56}$Fe,$^{58}$Ni and the whole mass region $63<\!A<\!90$
is reduced
and {\it s}-process yields are enhanced by $\sim$30\%.
The latter trend, which even amplifies the overproduction in this region 
compared to solar abundances,
stresses that 
in addition to the need for a better
assessment of
the rate of the neutron-producing reaction
$^{22}$Ne($\alpha$,n)$^{25}$Mg  \cite{jae:01s}, the
cross section data for other potential bottle neck nuclides 
({\it e.g.} $^{58}$Fe, $^{60}$Ni) must be
verified as well. 

Our direct measurement of the
Maxwellian-averaged ($kT\!=\!25$~keV) cross section of  
$^{62}$Ni($n,\gamma $)$^{63}$Ni
($28.4\pm 2.8$ mb) 
combines neutron activation with 
accelerator mass spectrometry, a technique which
complements the traditional experiments in cases
where the decay of the activation product cannot be
measured. The sensitivity
of nucleosynthesis calculations to individual cross sections
in the non-equilibrium situation of the weak
{\it s}-process, which significantly
contributes to the chemical history of the universe,
calls for an update of
the data set of ($n,\gamma $) astrophysical rates.

This work was supported in part 
by the US-DOE, Office of  Nuclear Physics, under Contract No.
W-31-109-ENG-38 and by the USA-Israel Binational 
Science Foundation (BSF).
A.H. and R.R. performed this work under the auspices of the 
U.S. Department of Energy at the Los Alamos National Laboratory
operated by the University of California under contract No.
W-7405-ENG-36.

\end{document}